\documentclass[english,aps,prl,superscriptaddress,twocolumn,showpacs,footinbib]{revtex4-1}
\usepackage{graphicx}  
\usepackage{dcolumn}   
\usepackage{bm}        
\usepackage{amssymb}   
\usepackage{babel}
\usepackage{verbatim}
\usepackage{amsmath}
\usepackage{setspace}
 \usepackage{epstopdf}
\graphicspath{{figures/}}
\usepackage{siunitx}
\begin{document}

\widetext

\title{Enhanced concentrations of nitrogen-vacancy centers in diamond through TEM irradiation}

\author{D. Farfurnik}
\affiliation{Racah Institute of Physics, The Center for Nanoscience and Nanotechnology, The Hebrew University of Jerusalem, Jerusalem 9190401, Israel}

\author{N. Alfasi}
\affiliation{Andrew and Erna Viterbi Department of Electrical Engineering, Technion, Haifa 32000, Israel}

\author{S. Masis}
\affiliation{Andrew and Erna Viterbi Department of Electrical Engineering, Technion, Haifa 32000, Israel}

\author{Y. Kauffmann}
\affiliation{Department of Materials Science and Engineering, Technion, Haifa 32000, Israel}

\author{E. Farchi}
\affiliation{Dept. of Applied Physics, Rachel and Selim School of Engineering, Hebrew University, Jerusalem 9190401, Israel}

\author{Y. Romach}
\affiliation{Racah Institute of Physics, The Center for Nanoscience and Nanotechnology, The Hebrew University of Jerusalem, Jerusalem 9190401, Israel}

\author{Y. Hovav}
\affiliation{Dept. of Applied Physics, Rachel and Selim School of Engineering, Hebrew University, Jerusalem 9190401, Israel}

\author{E. Buks}
\affiliation{Andrew and Erna Viterbi Department of Electrical Engineering, Technion, Haifa 32000, Israel}

\author{N.Bar-Gill}
\affiliation{Racah Institute of Physics, The Center for Nanoscience and Nanotechnology, The Hebrew University of Jerusalem, Jerusalem 9190401, Israel}
\affiliation{Dept. of Applied Physics, Rachel and Selim School of Engineering, Hebrew University, Jerusalem 9190401, Israel}


\date{\today}
\begin{abstract}
The studies of many-body dynamics of interacting spin ensembles, as well as quantum sensing in solid state systems, are often limited by the need for high spin concentrations, along with efficient decoupling of the spin ensemble from its environment. In particular, for an ensemble of nitrogen-vacancy (NV) centers in diamond, high conversion efficiencies between nitrogen (P1) defects and NV centers are essential, while maintaining long coherence times of an NV ensemble. In this work, we study the effect of electron irradiation on the conversion efficiency and the coherence time of various types of diamond samples with different initial nitrogen concentrations.  The samples were irradiated using a 200 keV transmission electron microscope (TEM). Our study reveals that the efficiency of NV creation strongly depends on the initial conversion efficiency as well as on the initial nitrogen concentration. The irradiation of the examined samples exhibits an order of magnitude improvement in the NV concentration (up to $\sim 10^{11}$ NV/$\SI{}{\centi\meter}^2$), without degradation in their coherence times of $\sim 180$ $\SI{}{\micro\second}$. We address the potential of this technique toward the study of many-body physics of NV ensembles and the creation of non-classical spin states for quantum sensing.
\end{abstract}

\pacs{76.30.Mi}
\maketitle

\paragraph{}
The study of quantum many-body spin physics in realistic solid-state platforms has been a long-standing goal in quantum and condensed-matter physics. In addition to the fundamental understanding of spin dynamics, such research could pave the way toward the demonstration of non-classical spin states, which will be useful for a variety of applications in quantum information and quantum sensing. One of the leading candidates for such studies is the negatively charged nitrogen-vacancy (NV) center in diamond, having unique spin and optical properties, which make it useful for various sensing applications \cite{Taylor2008, Balasubramanian2008, Pham2012, Mamin2014,Alfasi2016,Dolde2011,Wolf2015,Clevenson2015,Chen2017}, as well as a resource for quantum information processing and quantum simulation \cite{Cappellaro2009,Bennett2013,Weimer2013}. 
\paragraph{}
The current state-of-the-art is limited by the requirement of obtaining high spin concentrations while maintaining long coherence times. The sensitivity of magnetic sensing grows as the square-root of the number of spins \cite{Taylor2008,Pham2012}, thus enhanced NV concentrations could improve magnetometric sensitivities. Furthermore, enhanced NV concentrations could lead to strong NV-NV couplings, which together with long coherence times, achieved using a proper dynamical decoupling protocol \cite{Farfurnik2015}, could pave the way toward the study of many-body dynamics in the NV-NV interaction-dominated regime \cite{Cappellaro2009,Bennett2013,Weimer2013}. However, nitrogen defects not associated with vacancies (P1 centers) create randomly fluctuating magnetic fields that cause decoherence of the quantum state of the NV ensemble \cite{deSousa2009,BarGill2012}. As a result, in most cases it would be beneficial to increase the concentration of NV centers while keeping the nitrogen concentration constant, i.e. improve the N to NV conversion efficiency. 
\paragraph{}
A common technique for improving the conversion efficiency is the irradiation of the sample with electrons \cite{Schwartz2012,Kim2012,McLellan2016}, protons \cite{Botsoa2011}, or ions \cite{Aharonovich2009}, which creates vacancies in the lattice. Additional annealing mobilizes the vacancies, thus increasing their probability of occupying lattice sites adjacent to isolated nitrogens and forming stable NV centers. For example, NV concentrations of up to $\sim 45$ PPM were recently demonstrated using a highly impractical and specialized irradiation process, at an energy of 
2 MeV, a flux of $\sim 1.4\times 10^{13}$ e/$\SI{}{{\centi\meter^2}\second} $ with in-situ annealing at $700-800^oC$ for $285$ hours. This process resulted in NV-NV dipolar interactions with strength of $\sim 420$ $\SI{}{\kilo\hertz}$, contributing to the understanding of many-body spin depolarization dynamics \cite{Choi2017}. 
\paragraph{}
Here we demonstrate a practical and applicative irradiation process based on commonly available transmission electron microscopes (TEM), using standard samples used in the field. While this realistic scheme is expected to be limited in terms of the resulting NV-NV interaction strength,
with the use of proper dynamical decoupling protocols \cite{Farfurnik2015}, the NV-NV interaction-dominated regime could still be reached. Moreover, the resulting enhancement in the conversion efficiency could make NV ensemble sensors vastly smaller and therefore more practical for magnetic \cite{Taylor2008,Pham2012, Wolf2015}, thermal \cite{Clevenson2015}, and electric \cite{Chen2017} sensing. TEM electron irradiation can also be used to create spin-active defects in other solid-state systems, such as silicon carbide \cite{Bockstedte2008}.

Improved conversion efficiencies through TEM irradiation at $\sim 200$ keV were previously demonstrated in high-pressure-high-temperature (HPHT) \cite{Kim2012} and delta doped \cite{McLellan2016} samples.  It is necessary to extend these results, and study the effect of irradiation on samples that are more relevant to ongoing research: chemical vapor deposition (CVD), with as grown and implanted NVs, for which the improvement of conversion efficiencies is not trivial. In our work, we systematically study the effect of TEM irradiation on such samples. We achieve an order of magnitude improvement in the conversion efficiencies of implanted CVD samples, and analyze their contribution to magnetometry and many-body physics.   

\begin{figure}[!t]	
	\includegraphics[width=0.49\columnwidth]{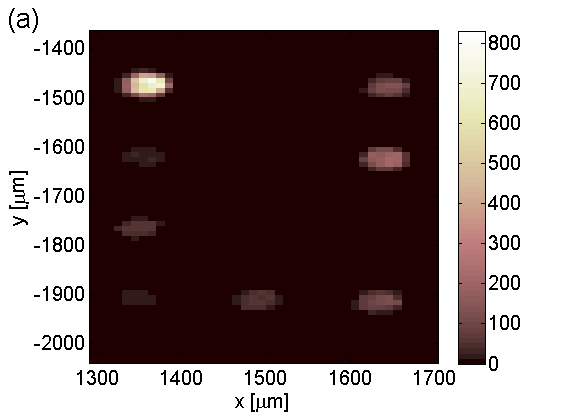} 
	\includegraphics[width=0.49\columnwidth]{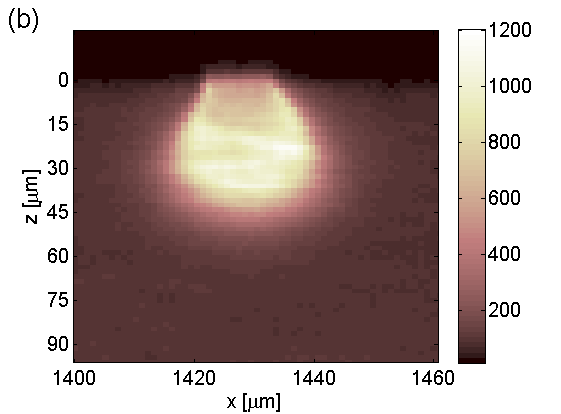} 
	\caption{(Color online) (a) Two dimensional (X-Y) scan of the surface of the HPHT sample, brightness level represents fluorescence produced by the NV centers in kilocounts per seconds. (b) Two dimensional (X-Z) scan of the fluorescence as a function of depth for a standard grade CVD sample, where $z=0$ represents the surface plane.}
	\label{fig:scanning}
\end{figure}
\begin{figure}[!t]	
	\includegraphics[width=0.88\columnwidth]{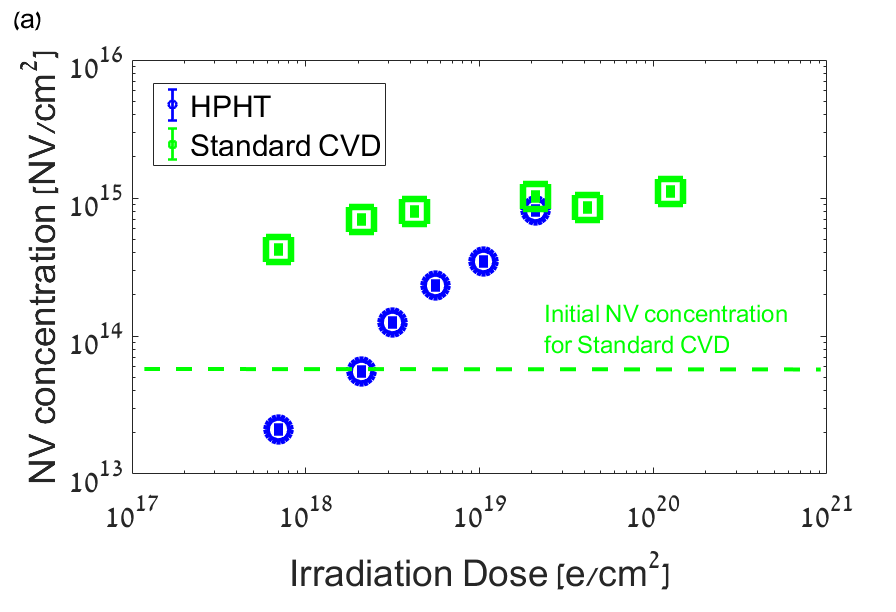} 
	\includegraphics[width=0.88\columnwidth]{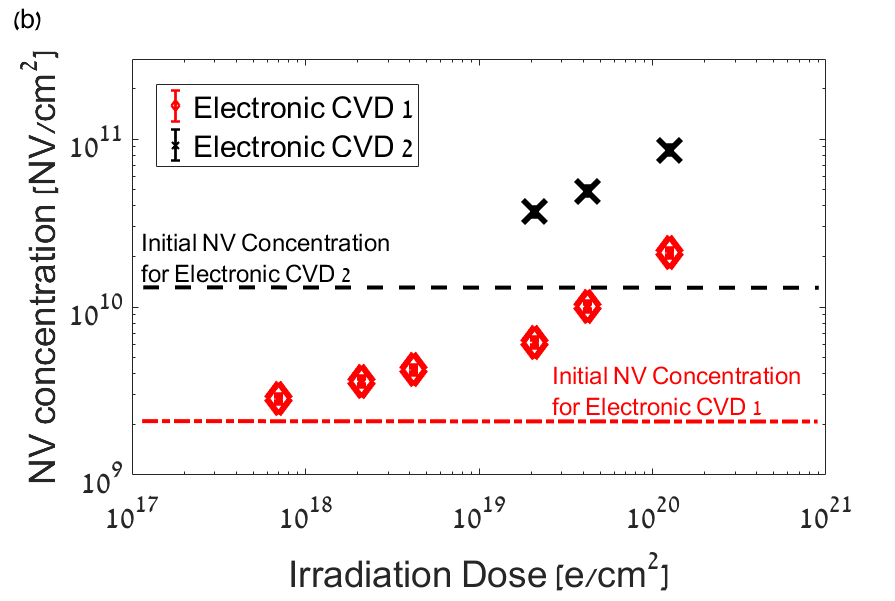} 
	\caption{(Color online) NV concentration as a function of the irradiation dose. (a) 3D samples - HPHT and standard grade CVD. (b) 2D samples - nitrogen implanted CVD. For clarity, the data points are enlarged with the error bars marked inside.}
	\label{fig:concentrations}
\end{figure}
\begin{figure}[!t]	
	\includegraphics[width=0.88\columnwidth]{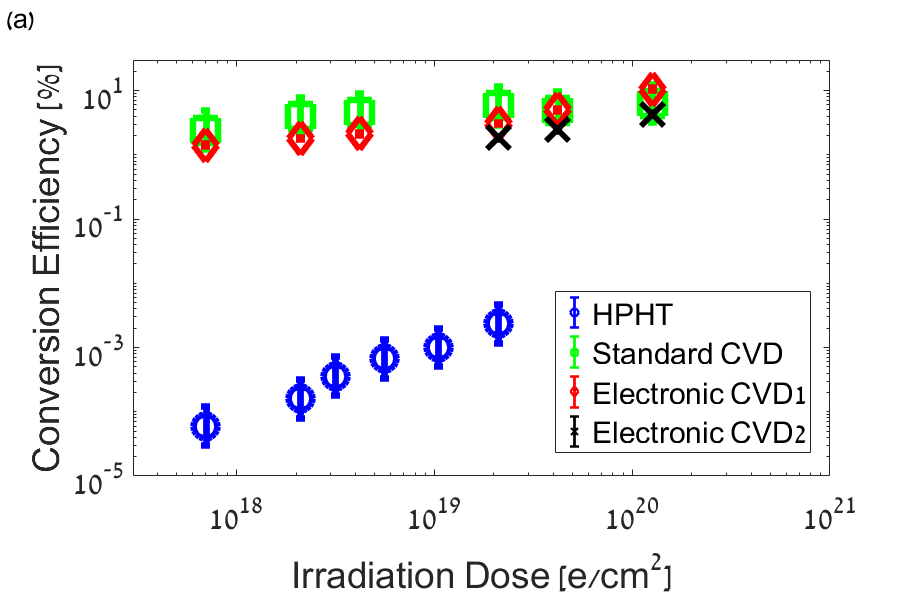} 
	\includegraphics[width=0.88\columnwidth]{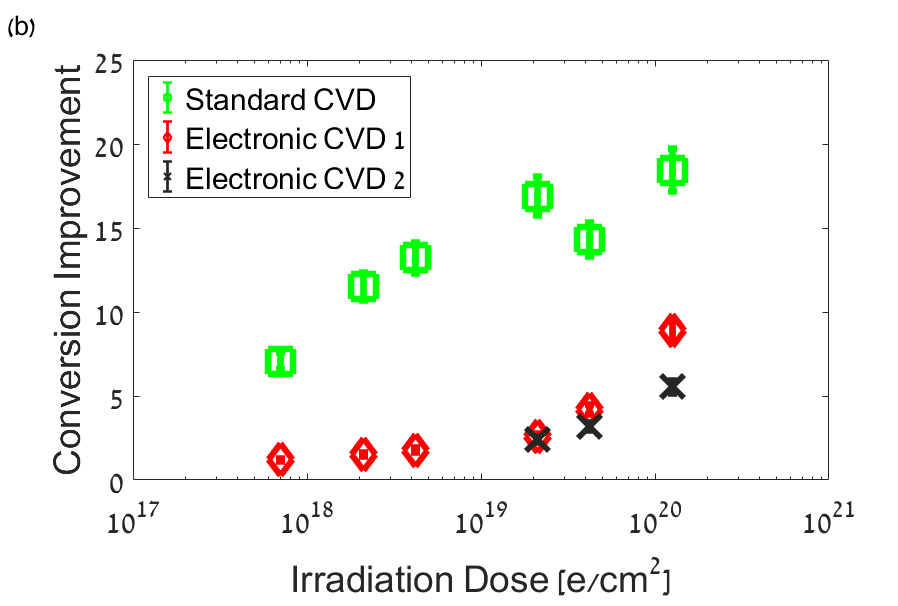} 
	\caption{(Color online)(a) N-to-NV Conversion efficiency as a function of the irradiation dose. (b) Improvement factor of the NV concentration due to the electron irradiation - the ratio between the NV concentration after irradiation, to the concentration before irradiation. Data for the HPHT sample is not shown since the initial NV concentration is negligible. For clarity, the data points are enlarged with the error bars marked inside.}
	\label{fig:efficiencies}
\end{figure}
\begin{figure}[!t]	
	\includegraphics[width=0.88\columnwidth]{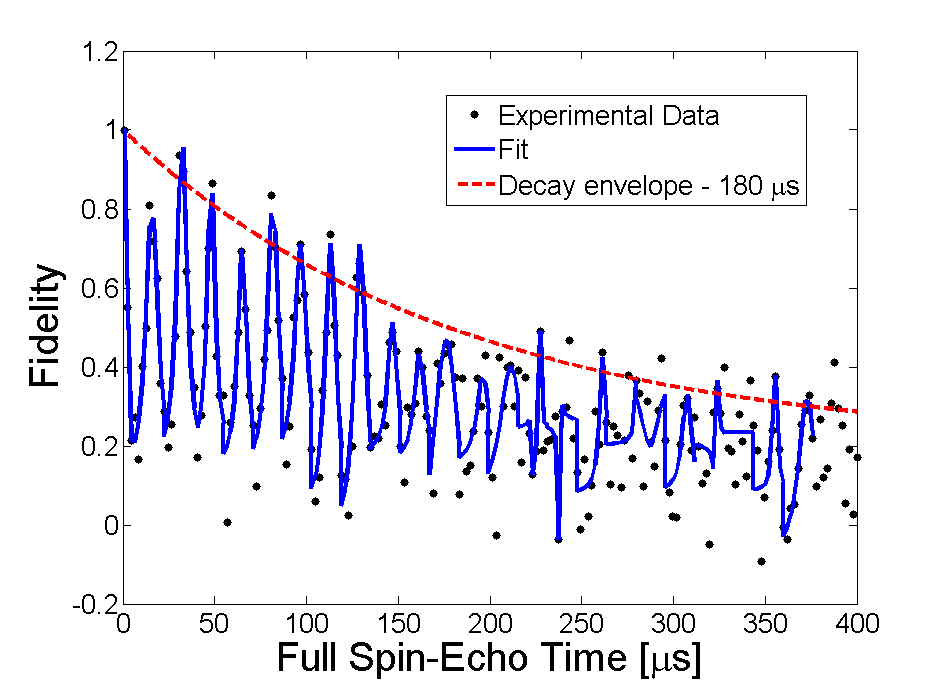} 
	\caption{(Color online) Hahn-Echo decoherence curve of an NV spin ensemble's state in a standard grade CVD sample, at an irradiation dose of $\sim7 \times 10^{17}$ e/$\SI{}{\centi\meter}^2$. The decoherence time is $T_2\approx 180$ \SI{}{\micro\second}. Revivals are caused by interactions with $^{13}C$ nuclear spins at a constant static magnetic field of $\sim 115$ G \cite{Gaebel2006}. The results for other irradiation doses are similar (see supplementary material).}
	\label{fig:echo}
\end{figure}

\paragraph{}
We study the effect of electron irradiation on four different samples (Element Six).
The first sample was produced by a standard HPHT technique, with an initial nitrogen concentration of $\sim 4 \times 10^{19}$ N/$\SI{}{\centi\meter}^3$ and poor conversion efficiency $(<10^{-5} \%)$. The second sample was produced by a standard CVD synthesis procedure, having a $\sim 2 \times 10^{16}$ N/$\SI{}{\centi\meter}^3$ initial nitrogen concentration (hereafter -  standard grade CVD). The last two samples were produced by a high purity CVD procedure with an initial nitrogen concentration of $\sim 2 \times 10^{14}$ N/$\SI{}{\centi\meter}^3$ and an NV concentration that was below the detection limit. These two samples then underwent a nitrogen implantation process (Innovion) \cite{Kalish1997,Meijer2005,Rabeau2006,Aharonovich2009,Toyli2010}, at an energy of 20 keV and doses of $2\times 10^{11}$ and $2\times 10^{12}$ N/$\SI{}{\centi\meter}^2$, followed by standard annealing (Across International TF1400): 8 hours, temperature $800^o$C, vacuum $\sim 7.5\times 10^{-7}$ Torr (hereafter - we refer to these two samples as nitrogen-implanted CVD). The maximum depth of the nitrogen layer created by the implantation is $\sim 100$ nm, limited by ion channeling \cite{Toyli2010}.  All samples were then irradiated using a 200 keV TEM (FEI Tecnai G2 T20 S-Twin) with doses ranging from $7.0\times 10^{17}$ to $1.3\times 10^{20}$  e/$\SI{}{\centi\meter}^2$, and experienced the same standard annealing. The diameters of the irradiated regions were $10-20$ \SI{}{\micro\meter}.  
\paragraph{}
We used a $532$ nm off-resonant laser in a home-built confocal microscope to induce fluorescence from NV centers. We located the irradiated regions by performing a two dimensional ($X-Y$) scan using precision piezoelectric translation stages (PI Micos LPS65) (Fig.  \ref{fig:scanning}(a)). Typically, the electrons create NV centers within dozens of microns inside the diamond (Fig. \ref{fig:scanning}(b)). The structure of the resulting NV-enhanced layer originates from multiple-scattering of electrons inside the diamond, consistent with previous Monte-Carlo simulations at similar irradiation conditions \cite{Kim2012}. 
\paragraph{}
In order to estimate the NV concentration, we used relatively low laser powers ($\sim 10$ \SI{}{\micro\watt}), for which the fluorescence signal is linearly proportional to the NV concentration. By comparing the signal to the fluorescence measured from a reference sample with a known NV concentration, we estimated the number of NVs within the measurement spot.  Dividing this number by the area / volume of the measurement spot yields the NV concentration per unit area / volume. The ratio between this concentration and the known nitrogen concentration gives the conversion efficiency.  First, we estimated the NV concentrations and conversion efficiencies in the non-irradiated regions (hereafter - ``initial" NV concentration / conversion efficiency), representing NVs that were present in the original lattice, or formed during the nitrogen implantation process. The resulting conversion efficiencies were $\sim 0.4\%$ for the standard grade CVD sample and  $\sim 1.2\%$ and $0.77\%$ for the nitrogen-implanted samples. Finally, the ratio between final and initial concentrations exhibits the factor of improvement in conversion efficiency solely by the TEM irradiation.
\paragraph{}
Since the NV concentrations are estimated by this fluorescence level, fluctuations in the fluorescence are the main source for errors. In our current setup, the dominant error source is a $\sim 3.5\%$ fluctuation in the laser power. The NV concentration is calculated by the ratio between the measured fluorescence level and the fluorescence level of a known sample, both having the same uncertainty, thus resulting in a total uncertainty in NV concentration of $\sqrt{2}\times 3.5\%$. For the calculation of the conversion efficiencies [Fig. 3(a)], NV concentrations are divided by the initial nitrogen concentrations, adding another source of even larger uncertainties: the nitrogen concentrations of the bulk samples were estimated by the manufacturer (Element Six). The manufacturer estimates the nitrogen concentration through a well-calibrated process and SIMS measurements, yielding values accurate to a factor of $\sim 2$. For the implanted samples, the dose is well known, with an uncertainty related to the statistics of the Poissonian implantation process. Although the resulting uncertainties in the absolute conversion efficiencies are relatively large due to the limited accuracy of estimating the nitrogen concetrations, these concentrations remain constant in all regions of a specific sample. For a specific sample, there is a clear correlation between the uncertainties of different data points for the conversion efficiencies, such that their improvement due to electron irradiation is apparent. In all of the experiments, the irradiation doses were directly measured at the TEM holder before inserting the sample, having negligible uncertainties compared to the above-mentioned uncertainties in the nitrogen and NV concentrations.
\paragraph{}
We plot the NV concentration results as a function of the irradiation dose for the bulk (Fig. \ref{fig:concentrations}(a)), and implanted (Fig. \ref{fig:concentrations}(b)) samples (full results are given in the supplementary material). On the one hand, the NV concentrations in the implanted samples are expressed per centimeter-squared. Since a nitrogen-implanted layer is much narrower than the depth of field in our measurements, the implanted samples could be treated as quasi two-dimensional samples for, e.g., vectorial imaging applications, for which the 2D concentration estimate is useful. On the other hand, the NV concentrations in the bulk samples are expressed per centimeter-cubed. In these samples, NV centers are excited from the whole measurement volume, making them more useful for, e.g., absolute AC/DC magnetometry, for which the 3D concentration estimate is useful. The irradiation process enhances the NV concentration by more than an order of magnitude, up to $\sim 10^{11}$ NV/$\SI{}{\centi\meter}^2$ for the implanted samples, and up to $\sim 10^{15}$ NV/$\SI{}{\centi\meter}^3$ for the 3D samples (which is, within our depth of field of $\sim 1\SI{}{\micro\meter}$, equivalent to a similar 2D concentration of $\sim 10^{11}$ NV/$\SI{}{\centi\meter}^2$). 
\paragraph{}
The achieved conversion efficiencies are close to $10\%$ for the implanted samples (Fig. \ref{fig:efficiencies}(a)). However, for most samples the NV concentration does not reach saturation, thus higher irradiation doses could lead to further enhancement of the conversion efficiency. In particular, due to its low initial conversion efficiency, the NV concentration in the HPHT sample grows with a very large slope. However, since the irradiation of the highest examined dose takes $\sim$ 30 minutes, the creation of much denser NV ensembles in HPHT samples is less practical. Moreover, since the nitrogen concentration remains high even after irradiation, the typical coherence times measured in HPHT samples are on the order of $\sim 5$ \SI{}{\micro\second} \cite{BarGill2012}, significantly reducing their potential for quantum sensing and the studies of many-body dynamics.
\begin{center}
	\begin{table*}[t]
		\small
		\begin{tabular}{| c | c | c | c | c | c | c | c | c |}
			\hline
			Sample & Init. N   & Init. conv. & TEM fin. & TEM fin. & $2.8$ MeV irrad.& $2.8$ MeV irrad.\\ 			
			Type & conc. & eff. $[\%]$ &  NV conc. & conv. eff. $[\%]$ & fin. NV conc. & fin. conv. eff. $[\%]$		
			\\ \hline
			HPHT &   $ 3.54^{+3.44}_{-1.77} \times 10^{19}$ [/$\SI{}{\centi\meter}^3$] & $<10^{-5}$ & $ 8.2^{+0.41}_{-0.41} \times 10^{14}$ [/$\SI{}{\centi\meter}^3$] & $0.0023^{+0.0023}_{-0.00115}$ & $4.3^{+0.21}_{-0.21} \times 10^{16}$ [/$\SI{}{\centi\meter}^3$]&  $0.12^{+0.12}_{-0.06}$ \\ \hline
			CVD &  $ 1.77^{+1.77}_{-0.885} \times 10^{16}$ [/$\SI{}{\centi\meter}^3$]  & $0.4^{+0.4}_{-0.2}$ & $ 1.1^{+0.054}_{-0.054} \times 10^{15}$ [/$\SI{}{\centi\meter}^3$]  & $6.3^{+6.3}_{-3.15}$ &$ 8.1^{+0.4}_{-0.4} \times 10^{14}$ [/$\SI{}{\centi\meter}^3]$ & $4^{+4}_{-2}$  \\ \hline
			imp. CVD 1 &  $2^{+4.5e-6}_{-4.5e-6} \times 10^{11}$  [/$\SI{}{\centi\meter}^2$]  & $1.2^{+0.06}_{-0.06}$ & $ 2.1^{+0.1}_{-0.1} \times 10^{10}$ [/$\SI{}{\centi\meter}^2$]  & $10.6^{+0.52}_{-0.52}$ & --- & --- \\ \hline
			imp. CVD 2 &  $2^{+1.4e-6}_{-1.4e-6} \times 10^{12}$  [/$\SI{}{\centi\meter}^2$]  & $0.77^{+0.038}_{-0.038}$ &$ 8.6^{+0.425}_{-0.425} \times 10^{10}$ /$\SI{}{\centi\meter}^2$] & $4.3^{+0.21}_{-0.21}$ & $ 5.7^{+0.28}_{-0.28} \times 10^{10}$ [/$\SI{}{\centi\meter}^2$] & $2.9^{+0.14}_{-0.14}$ \\
			\hline
		\end{tabular}
		\caption{NV concentrations and conversion efficiencies before and after irradiation. The presented results are for the highest examined dose for the TEM irradiation, and commerical irradiation process at an energy $\sim 2.8$ MeV and dose $\sim 8 \times 10^{17}$ e/$\SI{}{\centi\meter}^2$}
	\end{table*}
\end{center}
\paragraph{}
In Fig. \ref{fig:efficiencies} we analyze the data in terms of conversion efficiencies and their improvement following irradiation. First, for a sample with a lower initial conversion efficiency, the resulting enhancement is more significant. Assuming that a particular  irradiation dose creates a given number of vacancies, and the initial conversion efficiency is low, more isolated nitrogens are available for binding with the vacancies, thus forming NV centers. Since HPHT samples have a poor initial conversion efficiency, the irradiation process improves the NV concentration by more than two orders of magnitude. Similarly, due to its lower initial conversion efficiency, the standard grade CVD sample exhibits a higher improvement than the implanted sample, compared to the non-irradiated case (Fig. \ref{fig:efficiencies}(b)). Second, for a given initial conversion efficiency, the improvement factor depends on the initial nitrogen concentration: for the first 2D implanted sample having initial nitrogen concentration of $\sim 2 \times 10^{11}$ N/$\SI{}{\centi\meter}^2$, the NV concentration was improved by a larger factor ($\approx 9$) than the second 2D sample, having an initial nitrogen concentration of $\sim 2 \times 10^{12}$ N/$\SI{}{\centi\meter}^2$ (improvement factor of $\approx 5.5$), even though its initial conversion efficiency was higher ($1.22\%$ versus $0.77\%$). This effect is consistent with a vacancy limited process: the irradiation creates a given number of vacancies, and if vacancy concentration is the limiting factor in the binding process with nitrogens to form NVs, the conversion efficiency will be higher for a smaller initial nitrogen concentration.   
\paragraph{}
In order to take advantage of the sensing capabilities of the NV ensemble after irradiation, any arbitrary quantum state has to be preserved for a long coherence time. In Fig. \ref{fig:echo} we plot the decoherence versus time, for a Hahn-Echo experiment \cite{Hahn1950} performed on the standard grade CVD sample at a representative irradiation dose of $\sim7 \times 10^{17}$ e/$\SI{}{\centi\meter}^2$.  Within our measurement accuracy, even a coherence time as high as $\sim 180$ \SI{}{\micro\second} does not exhibit dependence on the irradiation dose (see supplementary material). The absolute fluorescence contrast drops with the irradiation dose, probably due to an increase in the steady-state fraction of NV0 defects \cite{Kim2012}. As a result, the chosen irradiation dose for NV applications should not exceed the level of saturation in the NV concentration (see supplementary material). The decoherence curve exhibits collapses and revivals, corresponding to the coupling of the NVs to surrounding $^{13}C$ nuclear spins, at their Larmor precession times \cite{Gaebel2006}. These dynamics limit applications in sensing and quantum many-body physics, and can be overcome by using isotopically pure $^{12}C$ samples. Compared to typical coherence times of  $\sim 5$ \SI{}{\micro\second} for HPHT samples and  $\sim 40$ \SI{}{\micro\second} for an implanted sample (see supplementary material), potentially limited by surface effects \cite{Romach2015}, the standard grade sample exhibits the longest coherence time, which is the most challenging to preserve.
\paragraph{}
Finally, we compare the conversion efficiencies achieved using the TEM irradiation, to those obtained by a commercially available high energy irradiation  process (Golan plastic, energy $\sim 2.8$ MeV, dose $\sim 8\times 10^{17}$ e/$\SI{}{\centi\meter}^2$) on samples with similar properties (summarized in Table I). Since the high energy irradiation is applied on a much larger area, small irradiation doses were available ($\sim 8\times 10^{17}$ e/$\SI{}{\centi\meter}^2$, two orders of magnitude smaller than the doses of the TEM). It is thus clearly seen (Table I) that except for the HPHT sample (with a low initial conversion efficiency), the resulting conversion efficiencies using the TEM irradiation are slightly higher than those achieved by the commercial process. As TEMs are available in many in-house nanotechnology facilities, they can be used as quick and efficient tools for the enhancement of NV concentration in high-quality diamond samples.  The locality of TEM-irradiation has major advantages, enabling the easy  creation of varying NV concentrations within a single sample. However, since TEM-irradiation of a whole centimeter-squared sample (which could be useful for magnetic imaging) is impractical, in these cases standard irradiation techniques are preferred. 
\paragraph{}
To summarize, we have shown that $200$ keV electron irradiation with doses up to $10^{20}$ e/$\SI{}{\centi\meter}^2$ can enhance the NV concentration of the examined samples by an order of magnitude, and reach conversion efficiencies of up to $10\%$ for nitrogen-implanted samples, with no degradation in their coherence properties. Since the NV concentration did not reach saturation, higher irradiation doses could lead to  further enhancement. The TEM irradiation could significantly improve the sensitivity of NV magnetometry, which grows as the square-root of the number of spins \cite{Taylor2008,Pham2012}. For example, a magnetometric sensitivity of $\sim 7$  $\frac{\SI{}{\nano\tesla}}{\sqrt{\SI{}{\hertz}}}$  was demonstrated using an NV ensemble in a sample with $1$ PPM nitrogen and $0.06\%$ conversion efficiency, measuring a $220$ KHz oscillating AC field \cite{Pham2012}. As demonstrated in Fig. \ref{fig:efficiencies}(a), the NV concentration can be increased by a factor of $\sim 10$, which can result in an improved sensitivity of $\sim 7/\sqrt{10}=2.2$ $\frac{\SI{}{\nano\tesla}}{\sqrt{\SI{}{\hertz}}}$. Furthermore, the achieved concentration of $\sim 10^{15}$ NV/$\SI{}{\centi\meter}^3$ corresponds to NV-NV dipolar coupling of $\sim 50$ $\SI{}{\hertz}$. By repeating the irradiation process on isotopically pure $^{12}C$ samples, where $\sim 30$ $\SI{}{\milli\second}$ coherence times can be achieved using optimized dynamical decoupling \cite{Farfurnik2015}, the NV-NV interaction dominated regime could be reached, opening an avenue for the study of many-body spin dynamics with available samples and processing techniques \cite{Cappellaro2009,Bennett2013,Weimer2013}.
\paragraph{}
\paragraph{}
See supplementary material for further information about decoherence, and full NV concentration and conversion efficiency results.
\paragraph{}
\paragraph{}

This work has been supported in part by the Minerva ARCHES award, the CIFAR-Azrieli global scholars program, the Israel Science Foundation (grant No. 750/14), the Ministry of Science and Technology, Israel, the Technion security research foundation, and the CAMBR fellowship for Nanoscience and Nanotechnology.

\bibliography{nvbibliography}

\end{document}